\documentclass[sigconf]{acmart}
\usepackage{xcolor}

\newif\ifshowrevisions
\showrevisionsfalse
\DeclareRobustCommand{\revise}[1]{\ifshowrevisions\textcolor{red}{#1}\else#1\fi}

\AtBeginDocument{%
  }

\setcopyright{none}
\renewcommand\footnotetextcopyrightpermission[1]{}

\begin{document}

\title{Living Inside the Black Box: Behavioral Probing and Adaptation in Mandatory Wearable Sensing}

\author{Yibo Meng}
\authornotemark[1]
\email{yim4007@med.cornell.edu}
\affiliation{%
  \institution{Weill Cornell Medicine, Cornell University}
  \city{New York}
  \state{New York}
  \country{USA}
}

\author{Bingyi Liu}
\affiliation{%
  \institution{University of Michigan}
  \city{Ann Arbor}
  \state{Michigan}
  \country{USA}}
\email{bingyi@umich.edu}

\author{Ruiqi Chen}
\affiliation{%
  \institution{University of Washington}
  \city{Seattle}
  \state{Washington}
  \country{USA}}
\email{ruiqich@uw.edu}

\author{Xiaolan Ding}
\affiliation{%
  \institution{North China University of Science and Technology}
  \city{Tangshan}
  \country{China}}

\author{Shuai Ma}
\affiliation{%
  \institution{Aalto University}
  \city{Helsinki}
  \country{Finland}}
\email{shuai.ma@aalto.fi}

\renewcommand{\shortauthors}{Meng et al.}

\begin{abstract}
Wearable sensing systems in high-stakes institutional contexts translate behavioral data into consequential judgments, yet wearers have little access to how those judgments are made. We present a qualitative study of 24 individuals who experienced mandatory electronic monitoring in China's community corrections system. We show that participants built what we term \emph{sensor literacy under constraint}, a practical form of risk-oriented knowledge developed through uncertainty, behavioral probing, and adaptation. We identify two orientations across rule domains. Where participants had mapped system behavior, they sometimes regained limited flexibility. Where uncertainty remained costly, they contracted movement and discretionary activity beyond formal rules. Some former wearers described residual habits of calculation after device removal. We discuss design implications for making institutional sensing intelligible to wearers, including sensor uncertainty, usable documentation, and evaluation after device wearing.
\end{abstract}



\maketitle

\section{Introduction}

Wearable sensing systems are increasingly deployed in high-stakes institutional contexts, where sensor data shape consequential decisions about people's lives \cite{weiser1999computer, abowd2000charting}. Electronic monitoring (EM) is an extreme instance of this condition. Individuals under community supervision wear GPS ankle devices, smart wristbands, or mobile applications continuously. These systems record location, movement, and behavioral data around the clock. The data are translated into compliance judgments with legal consequences. The interpretive logic behind those judgments remains largely inaccessible to the people wearing the devices.

\revise{In the deployment reported by participants, EM centered on location and geofence monitoring, scheduled check-ins, and device status such as battery level and signal availability. Wearers encountered the system mainly through a worn device, a paired application for some participants, and follow-up communication from supervising staff. We do not reconstruct operator-side settings or institutional decision rules. Instead, we analyze what was made legible to wearers and how they acted under those limits.}

Sensor records do not carry fixed meaning. A boundary crossing, signal interruption, or timing deviation may reflect GPS drift, environmental interference, or actual non-compliance \cite{lane2010survey, hightower2002location}. Wearers rarely know how such ambiguities are resolved. They must act not only under observation, but under uncertainty about how their actions will be evaluated \cite{rader2018explanations, kizilcec2016much}.

Prior work on EM has focused on compliance outcomes, recidivism, and institutional practices \cite{belur2020systematic}. \revise{It has also examined legal risks, false alerts, perceived punitiveness, and accountability around location-based supervision} \cite{malek2023legal, hwang2026false, Richter2021}. Research on wearable sensing in ubicomp and HCI has examined fitness trackers, health monitors, and other devices that users choose and can exit \cite{epstein2020, Consolvo2008}. These settings involve low stakes, user control, and voluntary participation. Much less is known about how people actively calibrate themselves to \emph{mandatory}, continuously worn sensing systems \revise{from the wearer side}. This includes how they figure out what the system records, what it tolerates, and what it does with data about them when the system offers little explanation and little practical exit.

To address this gap, we present a qualitative study of 24 individuals who experienced EM supervision in China's community corrections system, based on semi structured interviews. We treat EM as an analytically revealing case of wearable sensing. Devices may be worn 24 hours a day. Sensed data carry immediate legal implications. System operation remains opaque to wearers. This combination makes visible a dynamic that is present but harder to observe in other wearable contexts.

Our central argument is that people experimentally learn how to live under opaque wearable sensing systems. We term this process \emph{sensor literacy under constraint}. When wearers lack access to how their data are interpreted, they test the system in small ways. They stand near boundaries to watch the GPS dot, delay check-ins by a few minutes, and drain a device battery to measure its runtime. Over time, this learning produces orientations that vary by rule domain. In domains they had tested enough to trust, some wearers found limited working flexibility. In domains that remained uncertain, many narrowed movement and reduced discretionary activity. They stayed away from boundaries, returned home early, and avoided decisions that might brush against unknown rule edges. Some former wearers also described residual habits of caution after device removal.

This work makes the following contributions:
\begin{itemize}
    \item We document how wearers of mandatory sensing systems actively calibrate themselves to a system they cannot inspect, through what we term \emph{sensor literacy under constraint}. This is practical, bodily, and risk-oriented experimentation under institutional consequence. It extends prior work on folk theories by showing how probing changes when the system is continuously worn and difficult to exit.

    \item We characterize two orientations across rule domains. Participants reported limited flexibility where they trusted a local model of system behavior, and conservative contraction where uncertainty remained costly. Retrospective accounts from former wearers also describe residual habits of calculation after device removal.

    \item We derive three design implications for institutional wearable sensing systems: surfacing sensor uncertainty to wearers, providing wearer-facing documentation of system tolerances, and extending evaluation beyond the wearing period. Each implication is grounded in a specific gap between what wearers demonstrably needed and what the system provided.
\end{itemize}

EM is a high intensity case for examining wearable sensing that is difficult to refuse, consequential, remotely interpreted, and opaque to wearers. \revise{We do not claim that findings transfer directly from China's community corrections system to other settings. Instead, the case specifies a scope condition: the analysis is most relevant where wearable sensing is mandatory or hard to refuse, consequential, remotely interpreted, and opaque to wearers.} The mechanisms identified here \revise{may help analyze} workplace monitoring, insurance telematics, eldercare sensing, and rehabilitation tracking when those conditions are present. Understanding what these systems do to the people who wear them is a foundational question for wearable computing.

\section{Related Work}

\paragraph{Wearable sensing in everyday life.}
Ubiquitous computing research has long examined how sensing 
systems become embedded in daily routines and influence everyday 
behavior \cite{weiser1999computer, abowd2000charting}. Work on 
personal informatics and self-tracking has shown that wearable 
devices affect how people understand and regulate their own 
activities, from step counting to sleep monitoring 
\cite{epstein2020, epstein2015lived, Consolvo2008}. These studies reveal that 
sensing systems are not merely passive recorders but active 
shapers of daily life. However, this body of work largely 
concerns voluntary, user-initiated sensing in low-stakes 
contexts, where individuals retain substantial control over 
participation, data access, and interpretation. Electronic 
monitoring differs in fundamental ways: wearing is compulsory, 
data carry immediate legal consequences, and wearers have no 
access to how their sensor records are interpreted or acted 
upon \cite{belur2020systematic, Richter2021, Troshynski2008}. EM thus 
extends wearable sensing into a setting where sensing is 
institutionally enforced, consequential, and externally 
interpreted rather than personally managed. \revise{Our study extends this line of work by foregrounding how wearers themselves build practical knowledge when institutional sensing remains operationally opaque.}

\paragraph{Learning to navigate sensing systems.}
When sensing systems operate through mechanisms that are not directly visible, users construct working understandings from indirect cues. Research on algorithmic folk theories shows that people develop informal mental models to explain system outputs in domains such as social media ranking and content recommendation \cite{Eslami2016, DeVito2017}. These models are partial and uncertain, yet they guide how people interact with and adjust their behavior in response to algorithmic systems. Related work has shown that users actively probe system behavior, experimenting with actions, observing outcomes, and revising their understanding over time \cite{Eslami2016, Jhaver2024}. \revise{Surveillance and panopticism scholarship similarly shows that persistent observation can induce behavioral caution even without explicit knowledge of system rules} \cite{manokha2018}.

This prior work, however, has primarily examined episodic, low-stakes interactions. A person who dislikes how their Twitter feed works can delete the app. Getting a misleading restaurant recommendation has no legal consequence. The probing documented in folk theory research happens because users are curious or frustrated, and they can walk away. In mandatory wearable sensing, the situation is different. Wearers have little practical exit. A misunderstood record can trigger a hearing. Not knowing how the system works is itself a form of vulnerability. This changes what probing means, what it costs, and what it does to people over time.

\paragraph{Behavioral effects after worn systems.}
Wearable behavior change research has documented how continuous feedback from sensing devices shapes users' activity patterns, self-perception, and decision-making over time \cite{epstein2020, Consolvo2008}. These effects are generally studied as intended outcomes of device use, such as increased physical activity or improved health awareness \cite{epstein2015lived}. Less attention has been paid to behavioral residues of sensing systems that are imposed rather than chosen. When wearers lack access to how a worn system interprets their behavior, their activity can adjust toward risk reduction rather than ordinary goals. Continuous monitoring research, which shares with wearable sensing the feature of ongoing behavioral observation, suggests that sustained observation can reorganize behavior beyond single moments of use \cite{manokha2018}. We examine how this adaptation is reported through everyday bodily interaction with a continuously worn device \revise{\cite{zhao2025immersive,chen2025gestobrush}} and through former wearers' accounts after device removal.

\section{Methodology}

\subsection{Research Design}

This study uses qualitative interviews to investigate how individuals who experienced mandatory electronic monitoring calibrated themselves to a sensing system they had limited ability to inspect or exit. We focus on what people do in response to not knowing. We examine how they test the system, what they learn, how that learning shapes their behavior, and what traces they report after device removal.

\subsection{Participants}

\begin{table}[h!]
\centering
\caption{Participant demographics \revise{(U/R indicates urban/rural residence).}}
\label{tab:participants}
\small
\begin{tabular}{clllll}
\toprule
\textbf{ID} & \textbf{Age} & \textbf{Gender} & \textbf{U/R} & \textbf{Occupation} & \textbf{Education} \\
\midrule
P1  & 28 & M & R & Office Clerk        & Bachelor's \\
P2  & 32 & F & R & Courier             & High School \\
P3  & 35 & F & U & Construction Worker & Middle School \\
P4  & 55 & M & U & Chef                & High School \\
P5  & 29 & F & R & Unemployed          & High School \\
P6  & 34 & M & R & Package Courier     & High School \\
P7  & 26 & F & U & None                & High School \\
P8  & 33 & M & U & Office Staff        & College Degree \\
P9  & 30 & M & U & Street Vendor       & Middle School \\
P10 & 36 & M & R & Unemployed          & High School \\
P11 & 27 & M & U & College Student     & College Student \\
P12 & 31 & M & R & Office Clerk        & College Degree \\
P13 & 37 & F & R & Factory Worker      & Middle School \\
P14 & 33 & F & R & Retail Worker       & High School \\
P15 & 27 & M & U & Kitchen Worker      & High School \\
P16 & 45 & M & R & Company Employee    & Bachelor's \\
P17 & 29 & M & U & Ride-hailing Driver & High School \\
P18 & 44 & F & U & Unemployed          & Middle School \\
P19 & 35 & M & U & Construction Worker & Middle School \\
P20 & 30 & M & R & Delivery Rider      & High School \\
P21 & 32 & M & R & Unemployed          & High School \\
P22 & 31 & M & R & Banquet Worker      & High School \\
P23 & 34 & M & R & Informal Worker     & Middle School \\
P24 & 34 & F & U & Office Staff        & College Degree \\
\bottomrule
\end{tabular}
\end{table}

As shown in Table~\ref{tab:participants}, we recruited 24 participants. The sample included 16 men and 8 women. Ages ranged from 26 to 55, with a mean of 34. Recruitment used purposive and snowball sampling. Initial access was established through community corrections institutions, including local judicial offices, and the sample was subsequently expanded through participant referrals. All participants had experienced EM supervision for at least three months by the time of interview. At the time of interview, the sample included both current wearers and former wearers who reflected on the period after device removal. We use ``wearers'' to refer to people who had worn or were wearing an EM device. Claims about experiences after device removal are limited to participants who described that period. Monitoring durations ranged from three months to over two years.

Most participants wore GPS ankle devices for location and geofence monitoring. Six also used a paired mobile application, three wore smart wristbands, and two used additional physiological sensors. Because location monitoring was the common function across devices, our analysis focuses on location uncertainty and device management rather than comparing device models. \revise{Participants' access to the system was limited: they described seeing device or app feedback, receiving reminders or follow-up calls, and attending check-ins, but not seeing operator dashboards, tolerance rules, or the criteria used to classify ambiguous records.} Participants represented diverse occupations, education levels, and living contexts. Thirteen participants lived in rural contexts and eleven lived in urban contexts. After the 21st interview, additional interviews produced no major new codes related to uncertainty, probing, or adaptation.

In this setting, EM records are used in routine community corrections management. Location deviations, missed check ins, low battery, or signal interruptions can trigger follow up conversations, point deductions, warnings, or further review depending on the case and local practice. Participants therefore treated ambiguous records as consequential, since technical anomalies could become part of an institutional compliance judgment.

\subsection{Data Collection and Analysis}
Interviews were conducted one on one, lasted 45 to 65 minutes, were audio recorded with consent, and were transcribed verbatim. Interviews were conducted in Chinese. Quotations were translated into English by bilingual researchers and checked against the original transcripts to preserve meaning, especially for terms describing uncertainty, tolerance, and violation risk. The interview guide covered participants' day to day experience under monitoring, their understanding of system operation, how they learned what the system would and would not tolerate, behavioral strategies they developed, and any changes they noticed after device removal when applicable. We employed thematic analysis. Three researchers independently coded a shared subset of transcripts, then refined a shared coding framework through discussion. The framework was applied to the full dataset, with codes grouped into higher level themes. For example, uncertainty related codes covering GPS drift, boundary ambiguity, and unpredictable enforcement were consolidated under \emph{interpretive uncertainty}, and adjustment related codes under \emph{coping patterns}.

The research team consisted of scholars in human-computer interaction and social computing with no institutional affiliation to the corrections system. Participants were informed that the study had no connection to their supervision. The study received IRB approval from \revise{North China University of Science and Technology}. Participants provided written informed consent, were free to withdraw at any time, and received compensation of 50 RMB.

\section{Results}

We present findings organized around two themes:
how wearers actively calibrated themselves to a system 
they could not inspect, and what that calibration process 
did to their behavior during monitoring and, for some former 
wearers, after device removal. 
The starting condition for both was the same: not knowing 
how the system worked was not a temporary state. 
It persisted, and people had to act anyway.

\subsection{Calibrating to an Opaque System}

\subsubsection{Uncertainty as the Condition for Calibration}

All 24 participants described ongoing uncertainty about how 
their sensor data were interpreted. This is the condition 
that made calibration necessary. Unlike the anxiety of 
being watched by a person, this uncertainty concerned the 
gap between their actions and how those actions became data: 
which GPS positions would register as violations, how timing 
deviations would be evaluated, and which signal interruptions 
would trigger consequences. Behaving more carefully did not 
resolve it, because the problem was not their conduct but 
their lack of access to how the system interpreted it.

\begin{quote}
P16: ``Having my behavior inferred from data is more 
distressing than being directly watched by a person. When 
someone watches me, I know what they see. But electronic data 
is different. It turns me into a set of dots and lines, and 
then someone judges me as a person based on those dots and 
lines. I have no idea what they will think.''
\end{quote}

A recurring trigger for this uncertainty was discovering that sensor records did not match wearers' understanding of their own actions. This included arriving home on time but being flagged as late, or remaining near their front door but finding a boundary crossing in the record.

\begin{quote}
P22: ``Once I was just walking around outside my front door, 
nowhere near leaving, but later the record showed I had gone 
slightly outside. I had no idea what had happened. Later I 
gradually understood that the GPS signal drifts. But no one 
told me this. I found out by bumping into it myself.''
\end{quote}

This interpretive uncertainty had consequences that extended 
beyond the psychological. When wearers could not predict what 
the system would register as a violation, they also could not 
predict which movements and social connections were safe to 
maintain. Geofence boundaries did not merely constrain 
movement in the abstract. They severed physical connections 
to people and places that fell outside the permitted zone, 
often without warning or recourse.

\begin{quote}
P21: ``There was a step at the hospital entrance. I sat 
on that step and waited. My older brother came down every 
so often to update me: what she ate today, what the 
doctor said. I do not know how long I sat there.''
\end{quote}

The physical visibility of the device added a separate dimension of difficulty. Wearing a GPS ankle monitor in public meant continuous exposure to the judgments of others, independent of what the system was recording. In a separate context, P3 reflected on the cumulative weight of this exposure:

\begin{quote}
P3: ``Sometimes I think I would rather just be locked up. 
Wearing this on the outside makes it genuinely hard to 
live well, hard to face the strange looks from other 
people. It is truly painful. I strongly suspect that a 
very close friend of mine ended our relationship because 
of this.''
\end{quote}

\subsubsection{How Wearers Tested the System}

Uncertainty was not a static condition but one that wearers 
actively worked to reduce. When sensor records diverged from 
their understanding of their own actions, participants were 
compelled to revise their mental models of the system. Over 
time, many developed operational knowledge sufficient to 
guide daily decisions. This included an informal understanding of GPS drift margins, timing tolerances, and which behaviors were more likely to trigger alerts. All of it was constructed from personal experience rather than any official documentation.

\begin{quote}
P11: ``After using it for a while, I developed a sense of 
which areas were ambiguous and which were the real red lines. 
That sense was something I developed from living inside it.''
\end{quote}

Rather than stopping at this uncertain grasp, almost all 
participants described testing the system in deliberate, 
small-scale ways. Participants framed these tests as attempts 
to avoid accidental violations rather than evade supervision. 
Their goal was to establish a practical safety margin for 
ordinary compliant activity when formal rules did not specify 
how sensor records would be interpreted. They chose lower-stakes 
situations for these tests and supplemented what they learned 
with observations of other supervised individuals.

\textit{Spatial probing.} Participants stood near geofence 
boundaries to observe GPS drift behavior, testing the gap 
between physical position and recorded position.

\begin{quote}
P20: ``I stood at the boundary without moving, just watching 
whether the dot on my phone would drift outside. Sometimes 
it drifted out a little and then drifted back. I needed to 
know how large the drift range was, so I could know how 
large my truly safe activity space actually was.''
\end{quote}

\textit{Temporal probing.} Participants used lower-stakes 
check-in points to test the system's tolerance window, 
deliberately arriving early or late by small margins to 
observe whether deductions followed.

\begin{quote}
P17: ``I used a less important check-in time to experiment, 
arriving a few minutes early and then a few minutes late. 
I found that a few minutes' difference did not result in a 
deduction. But I did not dare use the important check-in 
point to test this. The cost would be too high.''
\end{quote}

\textit{Hardware probing.} Some participants also investigated the device itself: how long the battery lasted, under what conditions the signal dropped, and what happened when charge ran low.

\begin{quote}
P13: ``One day I did nothing else, just charged the ankle 
device to full and then waited for the battery to run out. 
I simply wanted to know how many hours it could last. That 
information was important. You cannot rely on guessing.''
\end{quote}

Participants were careful to calibrate the risk of each 
test, choosing lower-consequence scenarios for 
experimentation and supplementing personal tests with 
peer observation.

\begin{quote}
P5: ``I would not test everything myself. For some things, 
watching others is enough. Testing things yourself carries 
risk. Using other people's experience is less costly.''
\end{quote}

This peer observation was sometimes organized through 
informal networks among supervised individuals, who 
exchanged observations about system behavior at check-ins 
and other points of contact.

\begin{quote}
P11: ``Sometimes we run into each other during check-ins 
and chat for a bit. Who was recently called in for a talk, 
what new developments there have been somewhere else. This 
information is very useful, far more useful than what the 
official sources say. What the officials say is the rules. 
What we discuss is how the rules actually operate in 
practice.''
\end{quote}

One exception is notable: P21 described being too fearful 
to conduct any probing, suggesting that when monitoring 
instills extreme fear, wearers may forgo information 
gathering entirely and adopt maximum conservatism from the 
outset.

\begin{quote}
P21: ``I do not know. I do not dare to try. If something 
went wrong, I am afraid of receiving a heavier penalty.''
\end{quote}

Even for those who did probe, the knowledge gained 
resolved uncertainty at the level of action without 
eliminating it at the level of feeling.

\begin{quote}
P8: ``I already know roughly where that line is, but 
every time I approach it my heart still races. Knowing 
and feeling at ease are two different things.''
\end{quote}

What these behaviors share is that \revise{across spatial, temporal, hardware, peer-observation, and non-probing accounts (P20, P17, P13, P5, P11, P21),} wearers were building knowledge the system did not provide to them. They found out about GPS drift by bumping into unexpected records. They learned timing tolerances by running small experiments at lower stakes check-ins. They measured battery life by watching it drain. Participants reported receiving no wearer oriented documentation that made these tolerances usable in daily life. The knowledge had to be earned through the body, through time, and under real legal risk. We term this process \emph{sensor literacy under constraint}. Unlike voluntary forms of sensor literacy such as self tracking practices in personal informatics or folk theories about social media algorithms, sensor literacy under constraint is oriented toward risk management under a system that is externally interpreted and difficult to exit. That distinction changes what probing costs, what it feels like, and what it produces.

\subsection{Adaptation Across Rule Domains and Reported Residues}

\subsubsection{Two Orientations Across Rule Domains}

As participants' understanding of specific rule domains 
stabilized, their behavioral orientations varied by domain 
and perceived risk.

\textit{Strategic utilization} appeared in accounts from four participants (P11, P17, P18, P23). In rule domains where participants had established reliable mental models and assessed violation consequences as manageable, they sought to maximize flexibility within those boundaries. This meant mapping precise safe zones, setting proximity alerts, and building routines that preserved normal daily activity.

\begin{quote}
P17: ``I drew out the places I could go very clearly on 
the map, and set up a vibration alert on my phone when I 
was approaching the boundary. Within the safe zone I could 
live normally. My mind did not have to be running all the 
time.''
\end{quote}

Strategic utilization depended on experiential knowledge 
that was specific to a rule domain. Participants described 
this orientation only where they trusted their local model 
of system behavior.

\textit{Conservative contraction} appeared in accounts from 20 participants. It describes a pattern in which wearers restrict their behavior beyond what rules require. It differs from ordinary compliance in its excess. Participants did not simply follow rules. They shrank their activity space to avoid proximity to uncertainty. Where uncertainty persisted or violation consequences were perceived as severe, participants kept well away from boundaries rather than approaching them. They returned well before curfews rather than on time.

\begin{quote}
P19: ``I would rather take a detour than walk down that 
street near the boundary. It is not that the street is 
prohibited. I just do not want to be close to that line. 
What if the system drifts? Walking ten extra minutes is 
nothing, but those ten minutes give me peace of mind.''
\end{quote}

These orientations were not participant types. The same participant could act strategically in a mapped domain while remaining conservative in domains where system behavior remained uncertain. Among the 20 participants showing conservative contraction, three (P1, P4, P13) also reported behavior resembling strategic utilization in specific domains they had mapped.

\subsubsection{Reported Residues After Device Removal}

Participants also described cognitive and behavioral residues of monitoring. During monitoring, many described continuous calculation of location, remaining time, and device status.

Cognitively, this background calculation occupied attention regardless of activity.

\begin{quote}
P10: ``When I go out, my mind is simultaneously running 
several tracks: what time is it, how long before I have 
to be back, where am I right now, how far from the 
boundary, how much battery is left. When you want to do 
something else, there is always a feeling of something 
pulling at you from behind.''
\end{quote}

Over time, the burden of maintaining this calculation shaped how some participants interpreted institutional intent.

\begin{quote}
P22: ``Sometimes I wonder whether they make the rules 
this vague on purpose. The vaguer the rules, the less 
you dare to move. I do not know whether this thought 
is correct, but I genuinely feel this way.''
\end{quote}

Behaviorally, some former wearers reported that habits of calculating permitted locations and return times remained salient after device removal.

\begin{quote}
P14: ``I notice that now when I go out, I still 
instinctively think about where I can go today and when 
I need to be back. The device has been off for nearly 
three months.''
\end{quote}

\begin{quote}
P4: ``I notice I am now more hesitant about everything 
than I used to be. Before, if I wanted to go somewhere, 
I would just go. Now, even when there are no restrictions, 
I still think through many things first. That way of 
thinking seems to have grown into me. It is not about 
the device anymore, it is about everything.''
\end{quote}

These retrospective accounts identify reported residues of monitoring rather than measured psychological change. They show how habits of calculation and caution could remain salient for some participants after device removal.

\section{Discussion}

P20 stood at the edge of the permitted zone and watched the GPS dot on his phone. Sometimes it drifted outside the boundary line. Sometimes it drifted back. He stood there for a while, not moving, just watching. He was trying to understand how large his actual safe space was, not the space the rules described, but the space the device would record. Nobody had told him about GPS drift. He found it by standing still and paying attention.

This is, in miniature, what many participants did. They could not understand how their data were interpreted. They tested the system in small ways and built working knowledge from the results. Some of that knowledge was accurate. Some was approximate. All of it was self constructed because the system provided little usable information for wearers. Over time, participants who had mapped a rule domain could navigate that domain with some flexibility. In domains that remained uncertain, participants shrank behavior well inside what the rules required.

This pattern is what we call \emph{sensor literacy under constraint}. As experienced by participants, the system made data available for institutional interpretation while giving wearers little usable information about uncertainty, tolerances, or ambiguous records.

EM makes this dynamic visible at high intensity. \revise{Because our data come from China's community corrections system, the empirical claims should not be read as a general account of all EM deployments or all institutional wearables.} The relevant scope condition is the combination of mandatory or hard to refuse sensing, consequential interpretation, remote judgment, and limited information for the wearer. Under those conditions, calibration work can fall on the person being monitored. This lens is relevant to other institutional sensing settings when they share these features.

These findings distinguish sensing uncertainty from interpretive uncertainty. Wearers were not only uncertain about whether GPS points, check in times, or battery states were accurately recorded. They were also uncertain about how those records would be translated into institutional judgments. A brief boundary excursion, a late check in, or a signal interruption could be treated as technical noise, ordinary variation, or possible non-compliance. Because this interpretive layer was largely invisible to wearers, formal rules were not enough for daily action. Participants had to infer practical tolerances from experience. Sensor literacy under constraint therefore concerns more than learning sensor behavior. It is also a way of learning how sensed traces become consequential judgments.

\paragraph{Surface sensor uncertainty to wearers.}
A core source of anxiety in our findings was the inability to distinguish sensor error from genuine violation. As P16 described, being judged through sensor data felt more distressing than direct observation because data produces an account of behavior that the wearer cannot see or predict. GPS drift, signal interruptions, and timing edge cases all produced records that wearers could not interpret and could not contest. High-stakes wearable sensing systems should treat sensor uncertainty as information for wearers, not only for operators. In compliance settings, this \revise{should mean practical guidance for ordinary compliant action rather than detailed enforcement logic}. This includes visible accuracy estimates, clear battery and charging expectations, explanations for ambiguous alerts, and contestation channels. Participants who understood GPS drift described greater ability to manage daily movement. They usually reached that understanding through experimentation rather than disclosure. Prior work has shown that partial visibility into system logic can alter how users interact with automated systems \cite{kizilcec2016much} \revise{\cite{luo2025s}}. In high-stakes wearable contexts, our findings show how this visibility also matters for physical movement.

\paragraph{Design documentation for wearers, not only operators.}
The probing behaviors we observed arose from the absence of usable documentation for wearers. P13 spent an entire day charging the ankle device to full, then waiting for the battery to run out, timing how long it lasted. ``That information was important,'' she said. ``You cannot rely on guessing.'' Participants reported receiving no documentation that made battery runtime, geofence tolerance, or check in flexibility usable in daily life. High-stakes wearable systems should provide wearers with access to information they demonstrably need. This includes how location data are sampled, \revise{what day-to-day guidance is available for interpreting uncertain records,} and how ambiguous records are handled. When participants had to stand at boundaries, time check ins, and drain batteries to obtain this information, the learning burden moved from the system to the wearer. Documentation should also support evidence translation. Wearers need to know what kinds of contextual information can matter when records are ambiguous, such as timestamps, photos, mobility constraints \revise{\cite{su2025flymethrough}}, or environmental signal conditions. This would give wearers a way to understand how lived context can be connected to system records before a record becomes a violation. Consumer wearables routinely expose battery levels, GPS accuracy indicators, and sync status to their users. Institutional deployments should treat these forms of operational legibility as basic support for ordinary compliant use.

\paragraph{Account for behavioral consequences beyond the wearing period.}
As P4 described months after monitoring ended, the habit of thinking through what was permitted before going anywhere had grown into him. It was no longer about the device, but about everything. Other participants described related habits of calculating what was permitted and staying well inside boundaries that no longer existed. These retrospective accounts identify a pattern that current evaluation frameworks rarely capture \revise{\cite{zhang2025slideaudit}}. Most assessments of institutional sensing systems measure compliance during the monitoring period. Evaluation after device removal would ask how wearing a sensing system shapes ordinary movement, confidence, and decision-making afterward. As wearable sensing expands into workplaces, insurance, and healthcare \revise{\cite{chen2026knowledgecaremixedmethodsevaluation,meng2026engagement,meng2026tibetcpr}}, this question becomes relevant wherever systems are consequential, hard to refuse, and opaque to wearers \cite{murray2024chilling}. \revise{We treat these domains as analytically adjacent only under those conditions; they should not be assumed to reproduce the same dynamics as community corrections.}

\paragraph{Limitations.}
This study is based on cross sectional interviews in one institutional context \revise{in China}. The data capture participants' accounts of uncertainty, probing, and adaptation rather than system logs or clinical measures. Retrospective accounts after device removal identify reported residues, not trajectories over time. The sample includes supervised individuals but not supervising authorities or system operators \revise{or procurement and design records}. \revise{Therefore, we cannot infer institutional intent, compare wearer accounts with operator logs, or evaluate how specific interface choices were made.} Future work could examine how institutional practices and system interfaces produce the calibration asymmetry described here.

\section{Conclusion}

This study examined how individuals under mandatory electronic monitoring adapt to a continuously worn sensing system whose interpretive logic remains inaccessible to them. Our central finding is that people experimentally learn how to live under such systems through sensor literacy under constraint. This knowledge is built by testing the system rather than consulting it. It produces orientations that vary across rule domains. In mapped domains, some wearers achieve limited flexibility. In uncertain domains, many narrow movement and reduce discretionary activity beyond formal rules. Some former wearers also described residual habits of calculation after device removal. EM offers a high intensity case for wearable sensing that is consequential, hard to refuse, remotely interpreted, and opaque to wearers. Addressing this condition requires treating sensor uncertainty as information for wearers, providing documentation that makes system behavior legible, and evaluating how device wearing shapes conduct after removal. Understanding what these systems do to the people who live inside them is a foundational question for wearable computing.

\begin{acks}
\revise{We thank the participants who shared their experiences with us.}
\end{acks}

\bibliographystyle{ACM-Reference-Format}
\bibliography{sample-base}

\end{document}